\documentclass[11pt]{article}

   \usepackage{times}
\usepackage{cospar}
\usepackage[sectionbib]{natbib}
\renewcommand{\bibsection}{\section*{REFERENCES}}
\renewcommand{\bibhang}{\setlength{8mm}}
\renewcommand{\bibsep}{\setlength{0mm}}
\usepackage{fleqn,cospar}

\usepackage{url}

\usepackage{graphicx}
\usepackage[figuresright]{rotating}

  \newcommand{\GC}{Galactic Centre}
  \newcommand{\eg}{e.g. }
  \newcommand{\ie}{i.e. }
  \newcommand{\etal}{et al.}
  \newcommand{\SgrAEast}{Sgr~A East}
  \newcommand{\ASCA}{{\it ASCA}}
  
  \newcommand{\Chandra}{{\it Chandra}}
  \newcommand{\XMMN}{{\it XMM-Newton}}
  \newcommand{\XMM}{{\it XMM}}
\newcommand{\NH}{$N_{\rm H}$}
\newcommand{\NHUNIT}{H~cm$^{-2}$}
\newcommand{\FLUXUNIT}{erg~s$^{-1}$~cm$^{-2}$}

\title{SGR A EAST AND ITS SURROUNDINGS OBSERVED IN X-RAYS}

\author{M. Sakano\address{Department of Physics and Astronomy, University of Leicester,
        Leicester LE1 7RH, UK} \address{Japan Society for the Promotion of Science (JSPS)},
        R.S. Warwick$^{1}$,
        and
        A. Decourchelle\address{CEA/DSM/DAPNIA, Service d'Astrophysique, C.E. Saclay, 
        91191 Gif-sur-Yvette Cedex, France}}

\begin{document}

\maketitle

\begin{abstract}
We report the results of an XMM-Newton observation of Sgr A East and its 
surroundings. The X-ray spectrum of Sgr A East is well 
represented with a two-temperature plasma model with temperatures 
of $\sim$1 and $\sim$4~keV.  Only the iron abundance shows
clear spatial variation; it concentrates in the core of Sgr A East.  The
derived plasma parameters suggest that Sgr A East originated in a single
supernova.  Around Sgr A East, there is a broad distribution
of hard X-ray emission with a superimposed soft excess component 
extending away from the location of  Sgr A East both above and below the 
plane.  We discuss the nature of these structures as well as the close
vicinity of Sgr~A*.
\end{abstract}

\section*{INTRODUCTION}

The {\GC} region consists of many complex and interesting structures,
which can be observed in a variety of windows encompassing the radio to 
$\gamma$-ray bands ({\eg} Mezger
{\etal} 1996).  The brightest region in the radio continuum band is
named Sgr~A, which is resolved into Sgr~A West and {\SgrAEast}.  Sgr~A
West includes a super massive black hole Sgr~A$^*$, of which the mass is
(2--3)$\times 10^6$ M$_\odot$ ({\eg} Genzel {\etal} 2000), as well as a
three-arm spiral-like structure orbiting Sgr~A$^*$ (the
mini-spiral) and the central dense star cluster (IRS~16 star cluster).
Sgr~A West is surrounded by {\SgrAEast} in projection.  {\SgrAEast}
shows an oval shell-like structure in the radio
continuum.  There is quite good evidence for a physical interaction
between Sgr~A West and East ({\eg} Yusef-Zadeh {\etal} 2000), although
many of the details remain uncertain.

The shell-like structure of {\SgrAEast} suggests it may be a supernova
remnant (SNR), SNR 000.0+00.0 (Jones 1974).
On the other hand, alternative interpretations have also been
proposed, {\eg} the remnant of an explosion caused by the central
massive black hole Sgr~A$^*$.  In any case, both the origin of {\SgrAEast} 
and its past evolution remain open questions.

The surface brightness of the diffuse radio non-thermal emission
surrounding Sgr~A is much brighter than other nearby radio structures. 
Some of the latter are filamentary and possibly a part of another SNR 
({\eg} Ho {\etal} 1985). There are also giant molecular clouds 
in vicinity of Sgr~A,
which are
probably undergoing some physical interaction with the radio continuum
structures ({\eg} Coil and Ho 2000). The presence of
extended hot plasma as detected in X-ray observations ({\eg} Yusef-Zadeh 
{\etal} 2000; Koyama {\etal} 1996) must also play an important role 
in relation to both the structure and the evolution of Sgr~A.  
Furthermore, the detection of the 
electron-positron annihilation 
line, high-energy cosmic ray excess, and a GeV source
(Mayer-Hasselwander {\etal} 1998) from somewhere in the {\GC}
region, most likely related with Sgr~A$^*$ or {\SgrAEast}, implies highly
energetic activity and/or particle acceleration.

The ultimate goal of studies of the Sgr~A region is to understand the 
distribution and state of the matter, the complex nature of the 
interactions and the dominant emission and absorption processes 
in operation, as a step towards a better general understanding of the nature 
of the central regions of galaxies.  
At the same time, the more specific study of a SNR (candidate) in the 
{\GC} may help us to understand the special interstellar environment 
and hopefully provide information relating to its evolution and/or 
formation.

The hard X-ray imaging observations of the Sgr~A region were first made
with {\ASCA} (Koyama {\etal} 1996) and traced an extended distribution
of X-ray emission originating from a thin thermal plasma.
More recently this region has been observed by {\Chandra}/ACIS, at
a spatial resolution of 0.5 arcsec
(Baganoff {\etal} 2002; Maeda {\etal}
2002).  Sgr~A$^*$ was, for the first time, resolved in the X-ray band
(Baganoff {\etal} 2002) and, in addition, bright diffuse
X-ray emission from {\SgrAEast} detected (Maeda {\etal} 2002).  
The X-ray emission associated with {\SgrAEast} fills the inner part
of the region defined by a non-thermal radio shell and has a spectrum 
characteristic of thin thermal emission at a  temperature 
$kT\sim 2$ keV and a metal abundance $Z\sim 4$ (Maeda {\etal} 2002).

In this paper, we report the {\XMMN} results for {\SgrAEast} and its
surroundings as derived from  both X-ray image and spectral data.
{\XMMN} observations benefit from the large effective area and good
imaging capability ($\sim$5~arcsec) of the XMM mirrors as well as good 
energy resolution of the EPIC CCD detectors 
(${\Delta}E/E\sim0.02$ at 5.9~keV).  
With this capability, the quality of the spectrum and statistics of the 
image are remarkable. We adopt the
distance of 8.0~kpc to the {\GC} throughout this paper (Reid 1993).

\begin{figure}
\centering
\includegraphics[width=150mm]{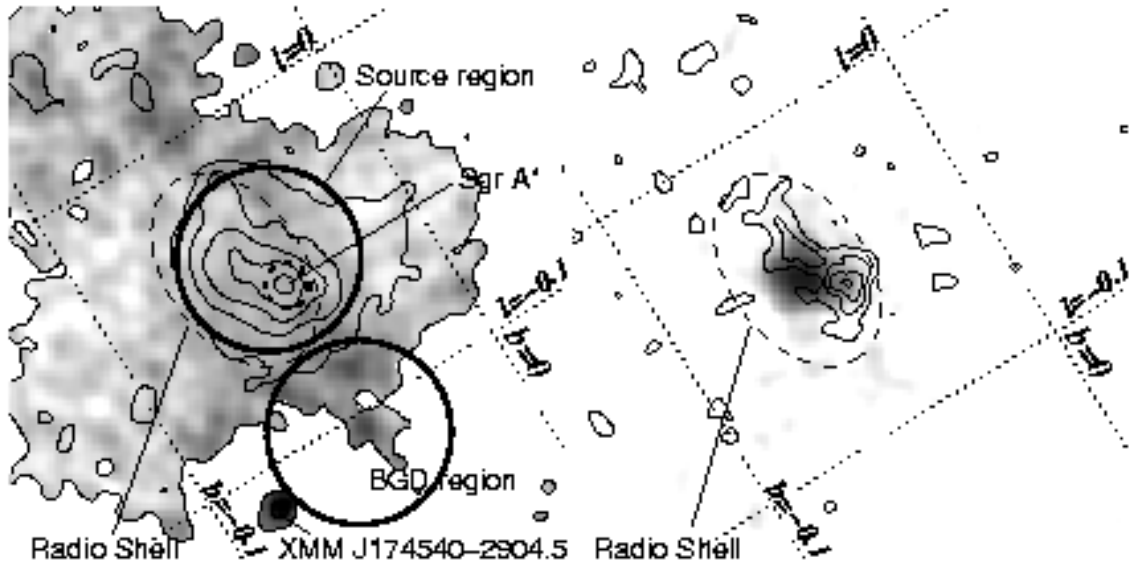}
 \caption{ \ \ {\sl Left:} 
The X-ray hardness ratio map (4.5--9~keV/2--4.5~keV) overlaid with the contours
of the 2--9 keV band image. Hardness ratio data are not plotted below the 
2--9 keV surface brightness threshold delineated by the lowest contour. 
Black represents harder areas in X-rays.
The spectrum accumulation region (solid circle: $r=100''$, dotted circle: $r=24''$,), the position of the radio shell and
the very hard source XMM~J174540$-$2904.5 are also indicated.
{\sl Right:} The He-like (6.7 keV) iron-line image overlaid with the He-like 
(2.4 keV) sulfur-line  contour.  The continuum is subtracted, assuming
the averaged spectral shape.  Whereas the sulfur-line contour is very
similar to that of the continuum, the iron-line image has a different
structure.  For both the image, the detector response, {\eg} the vignetting, 
is corrected.
  \label{fig:img}}
\end{figure}

\section*{OBSERVATION}

The {\XMMN} observation of the Sgr~A region was made on 2001 September 4.
In this paper, we concentrate on the results from the European Photon 
Imaging Camera (EPIC) on board {\XMMN}.  The EPIC instrument consists of two 
MOS CCDs cameras (MOS1 and MOS2) and one pn CCD camera,
which operate simultaneously.  The MOS and pn 
cameras were used in Full Frame and Extended Full Frame mode, respectively, 
with the medium filter selected.  Data reduction and filtering were carried
out with the Standard Analysis Software (SAS) Ver.5.4.
We accepted pixel patterns of 0--12 (single to quadruple events) for the
MOS CCDs but only single events (pattern 0) recorded by the pn system,
because of the calibration uncertainties for other pixel patterns in
the pn Extended Full Frame mode.
The effective exposure times for MOS1, 2 and pn after the background
filtering were 22.4~ks, 24.0~ks, and 17.5~ks, respectively.

\section*{X-RAY IMAGES OF SGR A EAST AND ITS SURROUNDINGS}

The left panel of Figure~\ref{fig:img} shows the X-ray hardness ratio
and the 2--9 keV surface brightness measured by the MOS (1+2) cameras 
within a $10' \times 10'$ field of view centred on Sgr~A.  The region 
surrounding Sgr~A$^*$, the central
massive black hole, is the brightest in the image, although Sgr~A$^*$
itself was not spatially resolved.
There is a further X-ray emission to the east from Sgr~A$^*$,
elongated nearly parallel to the
Galactic plane.  The e-folding radius of the core of this  extended source
is 28$''$ in the 2--10
keV band, although the region with significantly enhanced surface brightness
(relative to the surroundings) has a major axis dimension of
$\sim 200''$.

As shown in Figure~\ref{fig:img}, the X-ray emitting region is contained
within the radio shell and the molecular dust ring, as is also readily 
apparent from the {\Chandra} observation 
(Maeda {\etal} 2002).  This
suggests the strong correlation between 
the extended X-ray emission and the radio structure, {\SgrAEast}.
Henceforce, we call this X-ray emission {\SgrAEast}.

Figure~\ref{fig:img} also shows the hardness image in which  the regions 
to the north and south of {\SgrAEast} in Galactic coordinates
appear to exhibit enhanced soft X-ray emission. Alternatively, one  might 
argue that a narrow ($< 3'$ wide) strip along the Galactic plane  
has a harder spectral form.  These two options
suggest either the presence of a soft outflow from the {\GC}
or the preferential absorption of the soft (2--4.5 keV) X-rays
along the Galactic plane.  However, since the scale height of the 
molecular clouds in the {\GC} region is 5$'$--9$'$ (Tsuboi {\etal} 1999; 
Sakano 2000), which corresponds to the full extent of our X-ray picture
rather the narrow ``absorption'' feature identified above and  the fact that
the soft enhancement is significantly brighter than its adjacent regions,
favours the outflow argument. However, we note that the soft emission does 
not seem to emanate directly from Sgr~A$^*$, but from the region shifted by a 
couple of arcminutes east of Sgr~A$^*$.  Further detailed study is beyond the 
scope of this paper.

In the field of view, we discovered an apparently very hard source,
XMM~J174540$-$2904.5, $\sim 4'$ south of Sgr~A$^*$, as indicated 
in Figure~\ref{fig:img}.
This source was found to be extended and have a non-thermal radio
counterpart, which is well-aligned with the X-ray structure (Ho {\etal}
1985; Sakano {\etal} 2003). Detailed results on this source are presented 
elsewhere (Sakano {\etal} 2003).

Since the X-ray spectrum of {\SgrAEast} shows distinct emission lines
(see the next section), we also made line narrow-band images corresponding
to the He-like iron (6.7-keV) and sulfur (2.4-keV) lines. In this case
the underlying continuum is subtracted using adjacent bandpasses and 
assuming an averaged spectral shape (Figure~\ref{fig:img} right panel).
The 6.7-keV line is clearly more concentrated in the core of {\SgrAEast} 
than the continuum
(Figure~\ref{fig:img}).  This implies that the core of {\SgrAEast} is more
abundant in iron or higher in temperature, or perhaps 
a combination of these factors.  In contrast, the 2.4-keV line peak is
located on Sgr~A$^*$.   These features are quantitatively evaluated by 
applying spatially-resolved spectral analysis, as described below.

\section*{X-RAY SPECTRA OF SGR A EAST}

\begin{table*}
\vspace*{-10pt}
\begin{center}
  \caption[]{\ \ Ion temperature (keV) from each K-line. \label{tbl:line-kt}}
\begin{minipage}{\textwidth}
\centering
  \begin{tabular}{cccc}
   \hline
   S & Ar & Ca & Fe\\
   1.00$^{+0.08}_{-0.20}$ & 2.5$^{+0.2}_{-0.3}$ & 1.9$^{+0.7}_{-1.9}$ & 4.1$^{+0.2}_{-0.6}$\\
   \hline
 \end{tabular}
\end{minipage}
\end{center}
\end{table*}

\begin{figure}
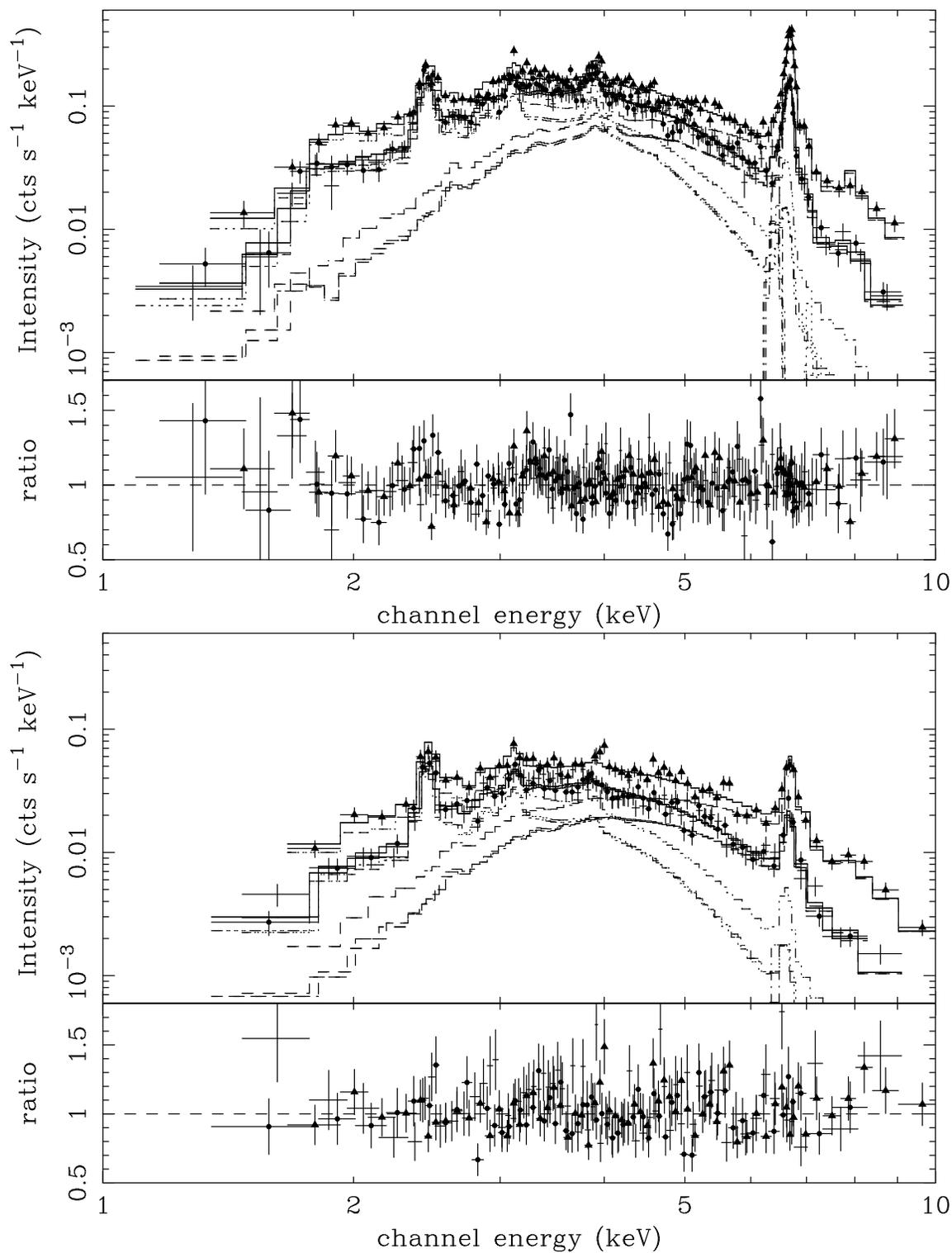

\begin{center}
 \includegraphics[width=100mm,angle=270]{sgraest_2vmkpcf_rat}
 \includegraphics[width=100mm,angle=270]{sgrastar_2vmkpcffit_rat}
\end{center}
 \caption{\ \ The simultaneous fitting of the spectrum of (top) Sgr A East 
(within a radius of 100$''$) and (bottom) Sgr~A$^*$ for pn (filled
 triangle), MOS1 (filled circle) and 2 (cross).  A two-component 
thermal plasma model is employed with metal abundances of Si, S,
Ar, Ca, and Fe allowed to vary.  The low energy spectrum is modified by a
partially covering absorber.  
 \label{fig:simul-fit}}
\end{figure}

We accumulated the source spectrum of Sgr A East from a region of
radius 100$''$ (see Figure~\ref{fig:img}), but excluding both Sgr~A$^*$ 
and its immediate surroundings
and the region around a bright soft point source.
We extracted a background spectrum from a nearby region at nearly
the same Galactic latitude so as to minimize the influence of the spatial
fluctuation of the background plasma.
Figure~\ref{fig:simul-fit} shows the resulting background-subtracted pn
spectra (although the spectral analysis was carried out using simultaneous
fits to the MOS 1, MOS 2 and pn spectra). Several strong
emission lines at energies corresponding to K$\alpha$ lines from 
highly ionized ions, can be seen.  This implies that a
significant amount of the emission originates in a hot thin thermal
plasma.

From the line ratios of K$\alpha$ lines from helium(He)-like and
hydrogen(H)-like atoms, we estimated the ionization temperature for each atom.
The temperatures are found to be
significantly different from element to element; $\sim$1~keV for sulfur,
$\sim$2~keV for argon and calcium, and $\sim$4~keV for iron
(Table~\ref{tbl:line-kt}).
The clear implication is that the plasma has a multi-temperature
characteristic (or originates in a combination of processes).

As a first approximation, we applied a two-temperature thin-thermal plasma 
model ({\sc MEKAL}) with a common absorption column, allowing the abundances 
of silicon, sulfur, argon, calcium, and iron in both the plasmas to be free.
The abundance of nickel was linked to that of iron.
In the best-fitting result (Table~\ref{tbl:fit-kt}) there still remains 
a counts excess below 2~keV.

Then, we examined the possibility that the absorption is patchy.
Specifically we assumed that a certain ratio $1-\epsilon$ ($\epsilon$: 
a free parameter) of the emission is absorbed by $7\times 
10^{22}${\NHUNIT} (fixed) with the remaining fraction subject 
to absorption by roughly twice the fixed column density (with the 
higher column density a free parameter of the fit).
This resulted in a significantly
improved fit (with a probability by the F-test of 99.97~\%;
(Table~\ref{tbl:fit-kt}; Figure~\ref{fig:img}) and
removed the bulk of the systematic trend apparent in the residuals.
The {\GC} region is known to be full of molecular clouds
of various sizes from radio and far infrared observations.  
A partial absorption model with a large covering fraction ({\ie}
small fraction for the `hole') may, therefore, be quite realistic.
We note that the inclusion of another lower temperature component
instead of a partial absorption model is found to give neither good
improvements in the fitting nor a realistic soft X-ray luminosity.

The main results of the spectral analysis are that: (i) the temperatures of 
the two components are $\sim$1~keV and $\sim$4~keV, which is fully 
consistent with the estimate based on the line-ratios; (ii) 
the lower-temperature component has by an emission measure an order of
magnitude larger than the higher-temperature component; (iii) the metal 
abundances are a few to several tens per cent higher than the solar value 
on average, although calcium is significantly more
abundant ($\sim$2 solar) than the other elements.

\begin{table*}
\vspace*{-10pt}
  \caption{\ \ The results from spectral fitting: \small
The unit for the normalisation is $10^{-12} \int n_{\rm e} n_{\rm H} dV / (4\pi D^2)$, where $n_{\rm e}$ and $n_{\rm H}$ are the electron and proton number densities (cm$^{-3}$) and $D$ is the distance of the source (cm).
The fluxes are for the 1--10 keV energy band.
The quoted uncertainties are at 90\% confidence for one interesting parameter.
\label{tbl:fit-kt}}
\begin{center}
\begin{minipage}{\textwidth}
\centering
 \begin{tabular}[tb]{ccccc}
 \hline
  Params. & Unit & \multicolumn{3}{c}{--- Data ---}\\
          &      & {\SgrAEast} & {\SgrAEast} & Sgr A$^*$\\
 \hline
Frac(i)   &          &  ---              & 0.93 (0.82--0.97) & 0.93 (fixed)\\
{\NH}(i) &({\NHUNIT})& 13.5 (12.8--14.2) & 14.5 (13.7--16.1) & 14.1 (13.2--15.0)\\
{\NH}(ii)&({\NHUNIT})&  ---              &  7.0 (fixed)      &  7.0 (fixed)\\
$kT_e$(1) & (keV)& 4.23 (3.82--4.71) & 4.40 (3.95--4.91) & 5.70 (4.75--6.79)\\
Norm(1)   &      & 1.66 (1.33--1.93) & 1.68 (1.40--2.02) & 0.47 (0.38--0.63)\\
$kT_e$(2) & (keV)& 0.94 (0.85--1.00) & 0.96 (0.90--1.03) & 0.96 (0.81--1.10)\\
Norm(2)   &      & 11.6 (9.5--12.7)  & 15.9 (11.5--20.3) & 4.0  (2.8--6.5)\\
$Z_{\rm Si}$   & & 5.1  (3.5--7.3)   & 1.83 (1.15--2.96) & 0.96 (0.36--1.94)\\
$Z_{\rm  S}$   & & 1.94 (1.56--2.47) & 1.46 (1.18--1.77) & 1.81 (1.41--2.34)\\
$Z_{\rm Ar}$   & & 1.23 (0.84--1.67) & 0.96 (0.67--1.27) & 1.12 (0.64--1.74)\\
$Z_{\rm Ca}$   & & 2.54 (2.04--3.07) & 2.04 (1.63--2.47) & 0.63 (0.05--1.23)\\
$Z_{\rm Fe,Ni}$& & 1.33 (1.18--1.51) & 1.26 (1.13--1.42) & 0.50 (0.40--0.63)\\
Fe-K$\alpha$ & ($10^{-5}$ph s$^{-1}$ cm$^{-2}$) & 1.4 (1.1--2.2) & 1.2 (0.8--1.8) & 0.1 (0.0--0.4)\\
 \hline
$\chi^2$/dof &   & 374.5/321 & 359.3/320 & 217.2/188\\
\multicolumn{2}{l}{$F_{\rm X}$ (MOS) ($10^{-12}$erg $s^{-1}$
  cm$^{-2}$)} & 12.6 & 12.5 & 3.4\\
 \hline
 \end{tabular}
\end{minipage}
\end{center}

\end{table*}

\subsection*{Spectral variation within Sgr A East \label{sec:separate-reg-fit}}

Next, we examine the possible spectral variation within {\SgrAEast}.
We extracted spectra from concentric regions of radii of 28$''$ (the
core region), 28$''$--60$''$ (the peripheral region) with
the centre at the peak of the 6.7-keV line image
(Figure~\ref{fig:img}), and from the outer part with a radius of
$>60''$ (the outer region) with the same centre as defined
in the previous sub-section (see Figure~\ref{fig:img}).  
We fitted these spectra with the
final version of the model described earlier, fixing the covering
fraction for the absorption at the best-fitting value.
Analysis shows that the iron abundance is significantly higher in the 
core (Z $\approx 3$) in comparison to the outer region (Z $\approx 0.5$).
In contrast the abundances of other metals do not vary significantly within
{\SgrAEast}.

\subsection*{Plasma parameters \label{sec:plasma}}

From the apparent extension of {\SgrAEast} in the hard X-ray band of
28~arcsec in radius, we calculate the total plasma volume
$V$ to be 1.6$\times 10^{56}$~cm$^3$, assuming a spherical shape.
The plasma is found to comprise two components and  we assume that
each component exists separately in the pressure balance with the other.  
Using the best-fitting parameters for the core region and
introducing the total filling factor $\eta_{\rm tot}$, which
is the sum of the filling factors of the two components, we estimated the
filling factor of $\eta_{\rm L}\approx 0.44 \eta_{\rm tot}$ and
$\eta_{\rm H}\approx 0.56 \eta_{\rm tot}$, the density of $n_{\rm
e,L}\approx 23 \eta_{\rm tot}^{-1/2}$ and $n_{\rm e,H}\approx 6.5
\eta_{\rm tot}^{-1/2}$, a total energy of $E\approx 1.5\times 10^{49}
\eta_{\rm tot}^{1/2}$ erg, and a total X-ray emitting mass of
1.8$\eta_{\rm tot}^{1/2}$ M$_{\odot}$, where the subscriptions of L and
H mean the lower- and higher-temperature components, respectively.

\section*{X-RAY SPECTRA OF SGR A*}

According to the results reported by Baganoff {\etal} (2002) 
based on the 1999 {\Chandra} observation, the compact X-ray source
in Sgr~A$^*$ is typically very faint with a flux $F_{\rm X}\sim
1.3\times 10^{-13}${\FLUXUNIT}, whereas the integrated
emission from other point sources and diffuse emission within a
10$''$-radius region is $\sim 1\times 10^{-12}${\FLUXUNIT}.  We did not
resolve Sgr~A$^*$ in our {\XMM} data (see the image section).  
We collected the spectrum from
a circular region of radius 24$''$, and found the flux to be
$\sim 3.4\times 10^{-12}${\FLUXUNIT}.  Taking the highly elevated
diffuse emission (Baganoff {\etal} 2002) into account, the integrated
flux measured with {\XMM}  are fully consistent with the {\Chandra} result.
Note the time interval including the Sgr~A$^*$ flare reported by
Goldwurm et al. (2003) was not included in our analysis.

We fitted the spectrum with the same model for {\SgrAEast}, with the
absorption parameters fixed.  Table~\ref{tbl:fit-kt} summarises the result.
We obtained temperatures of 0.96
(0.81--1.10) keV and 5.7 (4.8--6.8) keV, sulfur abundance of 1.8
(1.4--2.3), argon abundance of $\sim$1, and iron abundance
of 0.5 (0.4--0.6) solar.
As before,  the emission measure of the
lower-temperature component is about 10 times larger than that of the
higher-temperature one.  The increase in the higher temperature 
to $\sim$6~keV may be due to the contribution from discrete point sources,
including Sgr~A$^*$ itself.  We note that the iron abundance is found to be
significantly lower than {\SgrAEast}, $\sim$0.5 solar.  This low
iron-abundance is very similar to the value obtained for the
outer region of {\SgrAEast}.

\section*{DISCUSSION}

\subsection*{What is Sgr A East?}

The total energy of $\sim 1.5\times 10^{49} {\eta_{\rm tot}}$ erg is
smaller than the nominal energy for a SNR ($\sim 10^{51}$ erg).  Since
the plasma has already reached ionization equilibrium, this estimate
does apply to the full thermal energy in the observable X-ray band.
The total energy derived here is consistent with the {\Chandra} 
measurements reported by Maeda {\etal} (2002).
If {\SgrAEast} has a SNR origin, one SNR can easily account for
the observed structure and there is no need to invoke multiple 
supernovae as suggested by Mezger {\etal} (1989).

On the other hand, the estimated mass of 1.8${\eta_{\rm
tot}}^{\frac{1}{2}} {\rm M}_{\odot}$ is sufficiently large to account for a
supernova.  If there is some amount of unobserved mass, the total mass
may be even larger.  With an age of 8000 yr of {\SgrAEast} (Mezger
{\etal} 1989), this mass may originate as
either the ejecta or swept-up interstellar material.
The localisation of the iron abundance enhancement suggests that the mass
is predominantly that of the ejecta, and that a type-II SNR with a
low-mass progenitor or a type-Ia SNR is the likely scenario for the
origin of {\SgrAEast}.

The most remarkable characteristic of {\SgrAEast} is its unusually high 
temperature of 4~keV.  In fact, this temperature is
comparable with, or even higher than, the temperatures of young
`historical' SNRs, Cas~A, Kepler, and Tycho.
One possible scenario to explain this high
temperature is the shock has interacted with an ambient plasma already
preheated to a temperature of several keV.  In fact, it is suggested
that the {\GC} region is filled with the 10~keV plasma ({\eg} Koyama
{\etal} 1996), although other interpretations of the observed
X-ray spectra are now emerging (e.g. Dogiel {\etal} 2002).

\subsection*{Comments on the abundance in the Galactic Centre region}

The abundance in the {\GC} region is uncertain at present, mainly
because of the difficulty of the measurements.  The extrapolation of the
abundance gradient in the Galactic plane ({\eg} McWilliam {\etal} 1997)
points to an abundance higher than solar by a factor of 3 or more.  
A radio measurement by Mezger {\etal} (1979) gave an (oxygen)
abundance of $\sim$2~solar.  Recent measurements of
X-ray reflection nebulae model also give a relatively high (iron)
abundance for iron ($\sim 2$~solar) in the Sgr~B2 cloud (Murakami 
{\etal} 2001).  On the
other hand, measurements of the iron abundance in the atmosphere of infrared
stars in the {\GC} region, which may represent the most precise technique
available, give an inconsistent value of $\sim 1$
solar (Carr {\etal} 2000; Ram\'{\i}rez {\etal} 2000).

Our result shows that the iron abundance in the high-temperature
interstellar matter is sub-solar in 
the region around Sgr~A$^*$.  Therefore, it rather favours the smaller
abundance, suggested by the near infrared observations. We note that the
infrared star (IRS~7) observed by Carr {\etal} (2000) is located very close
(0.2~pc away) to Sgr~A$^*$, and thus samples basically the same region as 
the X-ray data.

\section*{ACKNOWLEDGEMENTS}

We are grateful to Dr. R. Willingale, Dr. Y. Maeda and Dr. S. Park for
 their valuable comments.  We thank Dr. R. Saxton, Dr. S. Sembay, Dr. G. 
 Griffiths, and Dr. I. Stewart for their comments and help on the
 {\XMMN} analysis software.  M. S. acknowledges the financial support
 from JSPS.

\medskip
\noindent E-mail address of M. Sakano\ \ \ {\underline mas@star.le.ac.uk}

\end{document}